\documentclass [12pt]{article}
\def\maj#1{\ifmmode\mbox{\usefont{U}{msb}{m}{n}#1}\else{\usefont{U}{msb}{m}{n}#1}\fi}
\def\v#1{\mathbf{#1}}
\makeatletter \@addtoreset{equation}{section} \makeatother

\usepackage{graphics,epsfig,graphicx}
\usepackage{color}

\begin{document}

\title{\textbf{How composite bosons really interact}}
\author{M. Combescot and O. Betbeder-Matibet
 \\ \small{\textit{Institut des NanoSciences de Paris,}}\\
\small{\textit{Universit\'e Pierre et Marie Curie and Universit\'e Denis
Diderot, CNRS,}}\\ \small{\textit{Campus Boucicaut, 140 rue de
Lourmel, 75015 Paris, France}}}
\date{}
\maketitle

\begin{abstract}
The aim of this paper is to clarify the conceptual difference which
exists between the interactions of composite bosons and the interactions
of elementary bosons. A special focus is made on the physical processes
which are missed when composite bosons are replaced by elementary bosons.
Although what is here said directly applies to excitons, it is also valid
for bosons in other fields than semiconductor physics. We in particular
explain how the two basic scatterings -- Coulomb and Pauli -- of our
many-body theory for composite excitons can be extended to a pair of
fermions which is not an Hamiltonian eigenstate -- as for example a
pair of trapped electrons, of current interest in quantum information.
\end{abstract}

			   
\newpage

In the 50's, theories have been developed to treat many-body effects
between quantum elementary particles, fermions or bosons, and their
representation in terms of Feynman diagrams has been quite enlightening
to grasp the physics involved in the various terms. This ``many-body
physics'' is now well explained in various textbooks [1-4].

While these theories have allowed a keen understanding of the microscopic
physics of electron systems, a fundamental problem remains up to now in
the case of bosons because essentially all particles called bosons are
composite particles made of an even number of fermions. Various attempts
have been made to get rid of the underlying fermionic nature of these
bosons, through procedures known as ``bosonizations'' [5]. By various
means, their main goal is to find a convincing way to trust the final
replacement of a pair of fermions --- for the simplest of these bosons
--- by an elementary boson, their fermionic nature being essentially
hidden in ``effective scatterings'', which supposedly take care of
possible exchanges between the fermions from which these composite bosons
are made.

A few years ago [6-8], we have decided to tackle the problem of
interacting composite bosons, with as a main goal, to find a way to
treat their interactions without replacing them by elementary bosons,
at any stage. It is clear that a many-body theory for composite bosons
is expected to be more complex than the one for elementary bosons.
However, here again, the diagrammatic representation we have recently
found [9], greatly helps to understand the processes involved in the
various terms, by making transparent the physics they contain.

The main difficulty with interacting composite particles made of fermions
is the concept of interaction itself. A first --- rather simple ---
problem is linked to the fact that, fermions being indistinguishable,
there is no way to know with which fermions these composite particles are
made. As a direct consequence, there is no way to identify the elementary
interactions between fermions which have to be assigned to interactions
between composite bosons: Indeed, if we consider two excitons made of two
electrons (e,e') and two holes (h,h'), there are six elementary Coulomb
interactions between them:
$V_{ee'},V_{hh'},V_{eh},V_{e'h'},V_{eh'}$ and $V_{e'h}$. While
$(V_{ee'}+V_{hh'})$ is unambiguously a part of the interaction between
the two excitons, $(V_{eh'}+V_{e'h})$ is the other part if we see the
excitons as made of $(e,h)$ and $(e',h')$, while this other part is
$(V_{eh}+V_{e'h'})$ if we see them as made of $(e,h')$ and $(e',h)$.
This in particular means that there is no clean way to transform the
interacting part of an Hamiltonian written in terms of fermions, into an
interaction between composite bosons. From a technical point of view,
this is dramatic, because, with an Hamiltonian not written as $H_0+V$,
all our background on interacting systems, which basically relies on
perturbation theory at finite or infinite order, has to be given up, so
that new procedures [10] have to be constructed from scratch, to
calculate the physiscal quantities at hand.

A second problem with composite bosons made of fermions, far more vicious
than the first one, is linked to Pauli exclusion between the boson
components. While Coulomb interaction, originally a $2\times 2$
interaction, produces many-body effects through correlation, Pauli
exclusion produces this ``$N$-body correlation'' at once, even in the
absence of any Coulomb process. In the case of many-body efects between
elementary fermions, this Pauli ``interaction'' is hidden in the
commutation rules for fermion operators, so that we do not see it. It is
however known to be crucial: Indeed, for a set of electrons, it is far
more important than Coulomb interaction, because it is responsible for
the electron kinetic energy which dominates Coulomb energy in the dense
limit. When composite bosons are replaced by elementary bosons, the
effect of Pauli exclusion is supposedly taken into account by introducing
a phenomenological ``filling factor'' which depends on density. In our
many-body theory for composite bosons, this Pauli exclusion appears in a
keen way through a dimensionless exchange scattering from which can be
constructed all possible exchanges between the composite bosons.

Since our many-body theory for composite bosons is rather new and not
well known yet, many people still thinking in terms of bosonized
particles with dressed interactions, it appears to us as useful to come
back to the concept of interaction for composite bosons, because it is
at the origin of essentially all the difficulties encountered with their
many-body effects, when one thinks in a conventional way, \emph{i}.\
\emph{e}., in terms of elementary particles. This in particular allows
to clarify the set of physical processes which are missed by any
bosonization procedure, whatever is the choice made for the effective
scatterings.

This paper is organized as follows:

In a first section, we briefly recall how elementary particles basically
interact and we give a few simple ideas on their many-body effects.

In a second section, we consider composite bosons made of two fermions,
\emph{a priori} different. We will call them ``electron'' and ``hole'',
having in mind, as a particular example, the case of semiconductor
excitons. We physically analyse what can be called ``interactions''
between two and between three of these composite bosons. We then show how
these physically relevant ``interactions'' can be associated to precise
mathematical quantities constructed from the microscopic
Hamiltonian written in terms of fermions.

In a third section, we discuss, on general grounds, the limits of what
can be done when composite bosons are replaced by elementary bosons
[11,12], in order to pick out which kind of processes are systematically
missed.

In a last section, we show a possible extention of the ideas of our
many-body theory for composite excitons to the case of composite bosons
which are not the exact eigenstates of the Hamiltonian, for example a
pair of trapped electrons, of current interest in quantum information
[13,14].

The goal of this paper is definitely not a precise application of our new
approach to any specific physical problem. In various previous
publications [10,15,16], we have already shown that our exact approach
produces terms which are missed when composite excitons are replaced by
elementary bosons with dressed interactions, these terms all having the
same physical origin. Since our approach now provides a clean and secure
way to reconsider problems dealing with semiconductor many-body effects
and optical nonlinearities --- through the virtual excitons to which the
photons are coupled ---, it appears to us as useful to clarify the
conceptual difference which exists between the possible  approaches to a
definitely difficult problem: many-body effects between composite
bosons, a problem of high current interest, in particular for its
consequences in BEC [17,18].

\section{Interaction between elementary bosons}

Let us call $|\bar{i}\rangle=\bar{B}_i^\dag|v\rangle$ a
one-elementary-boson state, its creation operator $\bar{B}_i^\dag$ being
such that
\begin{equation}
[\bar{B}_m,\bar{B}_i^\dag]=\delta_{m,i}\ .
\end{equation}
The concept of interaction between these elementary bosons is associated
to the idea that, if \emph{two} of them, initially in states $i$ and $j$,
enter a ``black box'', they have some chance to get out in different
states $m$ and $n$ (see fig.1a). In the ``black box'', one or more
interactions can take place (see figs.1b,c). However, since for
indistinguishable bosons, there is no way to know if the boson $i$
becomes
$m$ or $n$, the elementary process which can happen in the ``black box''
has to be the sum of the two processes shown in fig.1d.

From a mathematical point of view, this interaction between elementary
bosons appears through a potential in their Hamiltonian,
which reads
\begin{equation}
\bar{V}^{(2)}=\frac{1}{2}\sum_{mnij}\bar{\xi}_{mnij}^\mathrm{eff}\,\bar{B}_m^\dag
\bar{B}_n^\dag\bar{B}_i\bar{B}_j\ ,
\end{equation}
with
\begin{equation}
\bar{\xi}_{mnij}^\mathrm{eff}=\bar{\xi}_{nmij}^\mathrm{eff}\ ,
\end{equation}
due to the boson undistinguishability and
\begin{equation}
\bar{\xi}_{mnij}^\mathrm{eff}=\left(\bar{\xi}_{ijmn}^\mathrm{eff}\right)^\ast\
,
\end{equation}
due to the necessary hermiticity of the Hamiltonian.

To make a link between what will be said in the following on composite
bosons, it is interesting to note that, if the system
Hamiltonian
$\bar{H}$ reads $\bar{H}=\bar{H}_0+\bar{V}^{(2)}$, with $\bar{H}_0=\sum_i
\bar{E}_i\,\bar{B}_i^\dag\bar{B}_i$ and $\bar{V}^{(2)}$ given by
eq.(1.2), we have
\begin{equation}
[\bar{H},\bar{B}_i^\dag]=\bar{E}_i\,\bar{B}_i^\dag+\bar{V}_i^\dag\ ,
\end{equation}
with $\bar{V}_i^\dag|v\rangle=0$, while
\begin{equation}
[\bar{V}_i^\dag,\bar{B}_j^\dag]=\sum_{mn}\bar{\xi}_{mnij}^\mathrm{eff}\,
\bar{B}_m^\dag\bar{B}_n^\dag\ .
\end{equation}
This leads to an Hamiltonian matrix element in the two-boson subspace
given by
\begin{equation}
\langle v|\bar{B}_m\bar{B}_nH\bar{B}_i^\dag\bar{B}_j^\dag|v\rangle=
2[(E_i+E_j)\,\delta_{mnij}+\bar{\xi}_{mnij}^\mathrm{eff}]\ ,
\end{equation}
the scalar product of two-elementary-boson states being such that
\begin{equation}
\langle v|\bar{B}_m\bar{B}_n\bar{B}_i^\dag\bar{B}_j^\dag|v\rangle=
2\delta_{mnij}=\delta_{m,i}\delta_{n,j}+\delta_{m,j}\delta_{n,i}\ .
\end{equation}

If we now have three bosons entering the ``black box'', two interactions
at least are necessary, in order to find these bosons out of the box, all
three in a state different from the initial one (see figs.1e,f). Since 
$\bar{\xi}_{mnij}^\mathrm{eff}$ has the dimension of an energy, the
second scattering of this two-interaction process has to appear along
with an energy denominator.

\section{Interactions between composite bosons}

We now consider a composite boson made of two different fermions. Let us
call them ``electron'' and ``hole''. The case of composite bosons made of
a pair of identical fermions will be considered in the last part of this
work. We label the possible states of this composite boson by $i$.

\subsection{Two composite bosons}

We start by considering two composite bosons in states $i$ and $j$. From
a conceptual point of view, an ``interaction'' is a physical process
which allows to bring these bosons into two different states, $m$ and
$n$. What can possibly happen in the ``black box'' of fig.2a, to produce
such a state change?

\subsubsection{Pure carrier exchange}

The simplest process is, for sure, just a carrier exchange, either with
the holes as in fig.2b, or with the electrons as in fig.2c. Since the
two are physically similar, we expect them to appear equally in a
scattering $\lambda_{mnij}$ based on this pure exchange (see fig.2d).
It is of interest to note that the electron exchange of fig.2c is
equivalent to a hole exchange, with the $(m,n)$ states
permuted (see fig.2c').

If this carrier exchange is repeated, we see from fig.2e that two hole
exchanges reduce to an identity, \emph{i}.\ \emph{e}., no scattering at
all, while an electron exchange followed by a hole exchange
results in a $(m,n)$ permutation, \emph{i}.\ \emph{e}., again no
scattering at all for indistinguishable particles (see
fig.2f).

Let us now show how we can make appearing the
$\lambda_{mnij}$ exchange scattering formally. In view of fig.2d, this
scattering has to read
\begin{equation}
2\lambda_{mnij}=L_2\left(_m^n\ _i^j\right)+L_2\left(_n^m\ _i^j\right)\ ,
\end{equation}
where $L_2\left(_m^n\ _i^j\right)$ corresponds to the hole exchange of
fig.2b,
\begin{equation}
L_2\left(_m^n\ _i^j\right)=\int d\v r_e\,d\v r_h\,d\v r_{e'}\,d\v
r_{h'}
\,\langle n|\v r_{e'}\v r_h\rangle\langle m|\v r_e\v r_{h'}\rangle
\langle\v r_e\v r_h|i\rangle\langle\v r_{e'}\v r_{h'}|j\rangle\ ,
\end{equation}
$\langle\v r_e\v r_h|i\rangle$ being the wave function of the
one-boson state $|i\rangle$. Note that the prefactor 2 of eq.(2.1),
which could be included in the definition of the Pauli scattering, is
physically linked to the fact that \emph{two} exchanges are possible for
electron-hole pairs, namely an electron exchange and a hole exchange.
In the case of two electrons and one hole, as in problems dealing with
trions, these Pauli scatterings appear without any prefactor 2 because
an exciton can only exchange its electron with the electron gas.

If these one-boson states are orthogonal, $\langle
m|i\rangle=\delta_{m,i}$, it is tempting to introduce the
deviation-from-boson operator $D_{mi}$ defined as
\begin{equation}
D_{mi}=\delta_{m,i}-[B_m,B_i^\dag]\ ,
\end{equation}
where $B_i^\dag$ is the creation operator for the one-boson state
$|i\rangle=B_i^\dag|v\rangle$. For $\delta_{m,i}=\langle m|i\rangle$,
this operator is such that
\begin{equation}
D_{mi}|v\rangle=0\ ,
\end{equation}
while its commutator with another boson creation operator makes
appearing the exchange or Pauli scatterings, through
\begin{equation}
[D_{mi},B_j^\dag]=2\sum_n\lambda_{mnij}\,B_n^\dag\ ,
\end{equation}
as easy to see by calculating the scalar product of the two-boson states
$\langle v|B_mB_nB_i^\dag B_j^\dag|v\rangle$, using either the set of
commmutators (2.3,5) or the two-composite-boson wave function,
\begin{equation}
\langle\v r_{e'}\v r_{h'},\v r_e\v r_h|B_i^\dag B_j^\dag|v\rangle=
\frac{1}{2}\left[
\langle\v r_e\v r_h|i\rangle\langle\v r_{e'}\v r_{h'}|j\rangle
-\langle\v r_{e'}\v r_h|i\rangle\langle\v r_e\v r_{h'}|j\rangle\
+(i\leftrightarrow j)\right]\ .
\end{equation}
This wave function is indeed invariant by $(i\leftrightarrow j)$, as
imposed by $B_i^\dag B_j^\dag=B_j^\dag B_i^\dag$ for $B^\dag$'s being
products of fermion operators. It also changes sign under a $(\v r_e,\v
r_{e'})$ exchange, as required by Pauli exclusion.

This leads to
\begin{equation}
\langle v|B_mB_nB_i^\dag
B_j^\dag|v\rangle=2[\delta_{mnij}-\lambda_{mnij}]\ .
\end{equation}
This equation actually shows that the two-composite-boson states are
nonorthogonal. This is just a bare consequence of the fact that these
composite-boson states form an overcomplete basis [19]: Indeed, the
composite-boson creation operators $B_i^\dag$ are such that
\begin{equation}
B_i^\dag B_j^\dag=-\sum_{mn}\lambda_{mnij}\,B_m^\dag B_n^\dag\ ,
\end{equation}
easy to show by combining the fermion pairs in a different way.

Due to $B_i^\dag B_j^\dag=B_j^\dag B_i^\dag$,
eq.(2.7) also shows that
\begin{equation}
\lambda_{mnij}=\lambda_{mnji}=\lambda_{ijmn}^\ast\ .
\end{equation}
Finally, from the closure relation for one-boson states,
$\sum_i|i\rangle\langle i|=I$, it is easy to check that two exchanges
reduce to an identity, \emph{i}.\ \emph{e}., 
\begin{equation}
\sum_{rs}\lambda_{mnrs}\lambda_{rsij}=\delta_{mnij}\ ,
\end{equation}
with $\delta_{mnij}$ given in eq.(1.8), as physically expected from
figs.2e,f.

\subsubsection{Direct and exchange Coulomb scatterings}

If the two fermions are charged particles, another way for these two
composite bosons to interact is via Coulomb interaction between their
carriers. The simplest of these interactions is a set of direct
processes in which the out excitons $(m,n)$ are made with the same pair
as the ``in'' excitons $(i,j)$ (see figs.3a,b). However, here again, as
the carriers are indistinguishable, these processes must appear in a
scattering in which
$m$ and $n$ are not differentiated, as in fig.3c.

In view of figs.(3a,c), this direct Coulomb scattering must read
\begin{equation}
2\xi_{mnij}=C\left(_m^n\ _i^j\right)+C\left(_n^m\
_i^j\right)\ ,
\end{equation}
where, due to fig.3a, $C\left(_m^n\ _i^j\right)$ is given by
$$C\left(_m^n\ _i^j\right)=\int d\v r_e\,d\v r_h\,d\v r_{e'}\,d\v
r_{h'}\,\langle n|\v r_{e'}\v r_{h'}\rangle\langle m|\v r_e\v
r_h\rangle\, V(\v r_e\v r_h;\v r_{e'}\v r_{h'})\,\langle\v r_e\v
r_h|i\rangle\langle\v r_{e'}\v r_{h'}|j\rangle\ ,$$
\begin{equation}
V(\v r_e\v r_h;\v r_{e'}\v r_{h'})=V_{ee}(\v r_e,\v r_{e'})
+V_{hh}(\v r_h,\v r_{h'})+V_{eh}(\v r_e,\v r_{h'})+V_{eh}(\v r_{e'},\v
r_h)\ .
\end{equation}
The potential $V(\v r_e\v r_h;\v r_{e'}\v r_{h'})$ is just the sum of
the Coulomb interactions between an electron-hole pair made of $(e,h)$
and an electron-hole pair made of $(e',h')$. Note that, this
Coulomb scattering being \emph{direct}, the interactions are between
both, the ``in'' composite bosons $(i,j)$ and the ``out'' composite
bosons
$(m,n)$. From eqs.(2.11,12), we see that this direct Coulomb scattering
is such that
\begin{equation}
\xi_{mnij}=\xi_{nmij}=\left(
\xi_{ijmn}\right)^\ast\ .
\end{equation}

Let us now make appearing $\xi_{mnij}$ in a formal
way. If the one-boson states $|i\rangle$ are eigenstates of the
Hamiltonian, \emph{i}.\ \emph{e}., if
\begin{equation}
(H-E_i)\,B_i^\dag|v\rangle=0\ ,
\end{equation}
it is tempting to introduce the ``creation potential'' $V_i^\dag$
defined as
\begin{equation}
V_i^\dag=[H,B_i^\dag]-E_iB_i^\dag\ .
\end{equation}
Due to eq.(2.14), this operator is such that
\begin{equation}
V_i^\dag|v\rangle=0\ .
\end{equation}
If, as for the Pauli scattering $\lambda_{mnij}$, we consider the
commutator of this ``creation potential'' with another boson creation
operator, we can make appearing the direct Coulomb scatterings through

\begin{equation}
[V_i^\dag,B_j^\dag]=\sum_{mn}\xi_{mnij}\,B_m^\dag
B_n^\dag\ .
\end{equation}
The derivation of this result, without taking an explicit form of the
Hamiltonian, is however not as easy as the one for $\lambda_{mnij}$,
namely eq.(2.5), because, due to the overcompleteness of the
composite-boson states which follows from eq.(2.8), the
$\xi_{mnij}$ scattering of eq.(2.17) can as well be
replaced by $(-\xi_{mnij}^\mathrm{in})$, where $\xi_{mnij}^\mathrm{in}$
is an exchange Coulomb scattering defined as, (see fig.3d),
\begin{equation}
\xi_{mnij}^\mathrm{in}=\sum_{rs}\lambda_{mnrs}\,\xi_{rsij}\ .
\end{equation}
Consequently, this direct scattering $\xi_{mnij}$ cannot
be related to a precise matrix element as simply as for
$\lambda_{mnij}$ in eq.(2.7). Indeed, if we consider the matrix element
of the Hamiltonian $H$ between two-composite-boson states, we find,
depending if $H$ acts on the right or on the left,
\begin{eqnarray}
\langle v|B_mB_nHB_i^\dag B_j^\dag|v\rangle&=& 2[
(E_i+E_j)(\delta_{mnij}-\lambda_{mnij})+(\xi_{mnij}-
\xi_{mnij}^\mathrm{in})]\nonumber\\ &=& 2[(E_m+E_n)(\delta_{mnij}-
\lambda_{mnij})+(\xi_{mnij}-
\xi_{mnij}^\mathrm{out})]\ ,
\end{eqnarray}
where $\xi_{mnij}^\mathrm{out}$ is also an exchange Coulomb scattering,
this time defined as, (see fig.3e),
\begin{equation}
\xi_{mnij}^\mathrm{out}=\sum_{rs}\xi_{mnrs}\,\lambda_{rsij}\ .
\end{equation}
Due to eq.(2.19), these two exchange Coulomb scatterings,
$\xi^\mathrm{in}$ and $\xi^\mathrm{out}$, are linked by
\begin{equation}
\xi_{mnij}^\mathrm{in}-\xi_{mnij}^\mathrm{out}=(E_m+E_n-E_i-E_j)
\lambda_{mnij}\ ,
\end{equation}
while, due to eqs.(2.9,13), they are such that
\begin{equation}
\xi_{mnij}^\mathrm{in}=\xi_{nmij}^\mathrm{in}=\left(\xi_{ijmn}
^\mathrm{out}\right)^\ast.
\end{equation}

From the definitions of $\xi_{mnij}$ and $\lambda_{mnij}$ and the
closure relation for one-boson states, the
``in'' exchange scattering $\xi_{mnij}^\mathrm{in}$, shown in fig.3d, in
fact reads as
$\xi_{mnij}$ with $\langle n|\v r_{e'}\v r_{h'}\rangle\langle m|\v
r_e\v r_h\rangle$ replaced by
$\langle n|\v r_{e'}\v
r_h\rangle\langle m|\v r_e\v r_{h'}\rangle$. We see that $\xi_{mnij}
^\mathrm{in}$ contains
electron-hole Coulomb interactions which are \emph{between} the ``in''
states
$(i,j)$, but no more
\emph{between} the ``out'' states $(m,n)$ (see fig.3d').

In the same way, the ``out'' exchange
scattering $\xi_{mnij}^\mathrm{out}$, shown in fig.3e, reads as
$\xi_{mnij}$ with $\langle\v r_e\v r_h|i\rangle\langle\v r_{e'}\v r_{h'}
|j\rangle$ replaced by $\langle\v r_e\v r_{h'}|i\rangle\langle\v
r_{e'}\v r_h |j\rangle$; so that its electron-hole Coulomb interactions
are between the ``out'' states $(m,n)$ but no more between the ``in''
states
$(i,j)$.

$\xi_{mnij}^\mathrm{in}$ and $\xi_{mnij}^\mathrm{out}$ are Coulomb
scatterings with one exchange. If we now consider two exchanges, we can
think of them either on the same side as in fig.3f or on both sides as
in fig.3g. Two exchanges reducing to an identity, if these two
exchanges are on the same side, it is just the same as no exchange at
all. On the opposite, if they are on both sides, we end with something
very strange from a physical point of view. Indeed, the scattering
shown in fig.3g reads
\begin{eqnarray}
\int d\v r_e\,d\v r_h\,d\v r_{e'}\,d\v r_{h'}\,\langle n|\v r_{e'}\v
r_{h'}\rangle\langle m|\v r_e\v r_h\rangle\hspace{8cm}\nonumber
\\ \times\ [V_{ee}(\v r_e,\v
r_{e'}) +V_{hh}(\v r_h,\v r_{h'})+V_{eh}(\v r_e,\v r_h)+V_{eh}(\v
r_{e'},\v r_{h'})]\,\langle\v r_e\v r_h|i\rangle\langle\v r_{e'}
\v r_{h'}|j\rangle\ .
\end{eqnarray}
So that the electron-hole interactions $V_{eh}$ are not between the
composite bosons of any side. Being ``inside'' both composite bosons,
these
$V_{eh}$ interactions are already included in the composite bosons
themselves. Consequently, there is no physical reason for them to
appear once more in a scattering \emph{between} these composite
particles. This leads us to think that this type of exchange Coulomb
scattering should not appear in correct many-body calculations
involving composite bosons. And, indeed, we never produce them.

It is of importance to stress that there is only one physically
reasonable Coulomb scattering \emph{between} composite bosons, namely
$\xi_{mnij}$, because its electron-hole parts are unambiguously
interactions
\emph{between} the composite bosons on both sides. The proper way to see
the two other Coulomb scatterings, $\xi_{mnij}^\mathrm{in}$ and
$\xi_{mnij}^\mathrm{out}$, is in fact as a succession of a (direct)
Coulomb scattering before or after a carrier exchange. $\xi_{mnij}$ and
$\lambda_{mnij}$ actually form the two elementary blocks, necessary to
describe \emph{any} kind of interaction between composite bosons.
$\xi_{mnij}^\mathrm{in}$ and $\xi_{mnij}^\mathrm{out}$ are just two
among many other possible combinations of the two elementary blocks.
This is going to become even more transparent for the interactions
between three composite bosons.

\subsection{Three composite bosons}

We now consider what can be called interaction in the case of three
composite bosons, \emph{i}.\ \emph{e}., what physical processes can
transform the composite bosons $(i,j,k)$ into the composite bosons
$(m,n,p)$ (see fig.4a). If there is no common state between $(i,j,k)$
and
$(m,n,p)$, all three composite bosons have to be ``touched'' in some
way by this interaction, in order to change state.

\subsubsection{Pure carrier exchange}

As for two composite bosons, the simplest ``interaction'' between three
composite bosons is surely a carrier exchange. A possible one is shown
in fig.4b, with some of its equivalent representations shown in
figs.4c,d: It is easy to check that, in these three diagrams, the
composite boson
$p$ is made with the same electron as $j$ and the same hole as $k$. 

We can think of drawing diagram (4b) with the electron/hole lines
exchanged. As shown in fig.4e, this is however equivalent to a
permutation of the boson indices: Indeed, in the two diagrams of this
figure, the
$m$ boson has the same electron as $j$ and the same hole as $i$.

It is also of interest to note that the three-body ``skeleton diagram''
of fig.4b can actually be decomposed, in various ways, into exchanges
between two composite bosons: Indeed, diagram (4c) can be drawn as (4f)
and diagram (4d) as (4g), so that
\begin{equation}
L_3\left(\begin{array}{lll}p&k\\n&j\\m&i\end{array}\right)=
\sum_r\
L_2\left(\begin{array}{ll}n&k\\p&r\end{array}\right)\,
L_2\left(\begin{array}{ll}r&j\\m&i\end{array}\right)=\sum_s\ 
L_2\left(\begin{array}{ll}n&s\\m&i\end{array}\right)\,
L_2\left(\begin{array}{ll}p&j\\s&k\end{array}\right)
\end{equation}

Since the composite bosons are made with indistinguishable particles,
such a three-body exchange must however appear in a symmetrical way
through a scattering $\lambda_{mnpijk}$ which must read
\begin{equation}
3!\,2!\,\lambda_{mnpijk}=L_3\left(\begin{array}{lll}p&k\\n&j\\m&i
\end{array}\right)\ +11\  \mathrm{similar\ terms}\ ,
\end{equation}
obtained by permutating $(m,n,p)$ and $(i,j,k)$ (see fig.4h), all the
other positions of $(m,n,p)$ and $(i,j,k)$ being topologically
equivalent to one of these 3!2! terms. On that respect, it is of
interest to note that the factor of 2, in the definition (2.1) of the
Pauli scattering between two composite bosons $\lambda_{mnij}$, is
just 2!1!.  Due to fig.4b, the elementary exchange between three
composite bosons simply reads
\begin{equation}
L_3\left(\begin{array}{lll}p&k\\n&j\\m&i\end{array}\right)=
\int d\{d\v r\}\,\langle p|\v r_{e'}\v r_{h''}\rangle
\langle n|\v r_{e''}\v r_h\rangle\langle m|\v r_e\v r_{h'}\rangle
\langle\v r_e\v r_h|i\rangle\langle\v r_{e'}\v r_{h'}|j\rangle
\langle\v r_{e''}\v r_{h''}|k\rangle\ .
\end{equation} 

This three-body Pauli scattering $\lambda_{mnpijk}$ in particular
appears in the scalar product of three-composite-boson states,
\begin{equation}
\langle v|B_mB_nB_pB_i^\dag B_j^\dag B_k^\dag|v\rangle=\delta_{mnpijk}
-2(\delta_{m,i}\lambda_{npjk}\ +8\ 
\mathrm{permutations})+12\lambda_{mnpijk}\ ,
\end{equation}
with $\delta_{mnpijk}=\delta_{m,i}\delta_{n,j}\delta_{p,k}$ +5\ 
permutations, as possible to check either directly from the explicit
value of the composite boson wave function, or by using a commutator
technique based on eqs.(2.3,5) and on
\begin{equation}
3\lambda_{mnpijk}=\sum_r[\lambda_{mnri}\lambda_{prjk}
+\lambda_{mnrj}\lambda_{prik}+\lambda_{mnrk}\lambda_{prij}]\ ,
\end{equation}
which makes use of eq.(2.24).

\subsubsection{One Coulomb scattering}

If we now consider processes with one Coulomb scattering
only, it is necessary to have one additional exchange process at
least, to possibly ``touch'' the three composite bosons: See for
example the process of fig.5a, which precisely reads
\begin{eqnarray}
\sum_sL_2\left(_n^p\ _s^k\right)\,C\left(_m^s\ _i^j\right)=
\int \{d\v r\}\,\langle p|\v r_{e''}\v r_{h'}\rangle\langle n|\v r_{e'}
\v r_{h''}\rangle\langle m|\v r_e\v r_h\rangle\hspace{3cm}\nonumber\\
\times\ V(\v r_e\v r_h;\v r_{e'}\v r_{h'})\,\langle\v r_e\v r_h|i\rangle
\langle\v r_{e'}\v r_{h'}|j\rangle\langle\v r_{e''}\v r_{h''}|k\rangle\
.
\end{eqnarray}

Of course, we can also have one Coulomb and two exchanges, as obtained
by adding one Coulomb interaction wavy line in the three-body skeleton
diagram of fig.4b (see fig.5b): In the process of fig.5b, the ``out''
bosons are all constructed in a different way, while in the one of
fig.5a, one composite boson, among the three, stays made with the same
fermions.

\subsubsection{Two Coulomb scatterings}

Finally, as in the case of elementary bosons, it is also possible to
``touch'' the three bosons $(i,j,k)$ by two direct Coulomb processes,
as in fig.6a. Of course, additional fermion exchanges can take
place, if the ``in'' and ``out'' bosons are made with different pairs.
From a topological point of view, the processes in which the three
``out'' bosons are made with different pairs can be constructed from
the skeleton diagram of fig.4b, with the two direct Coulomb scatterings
being \emph{a priori} at any place, \emph{i}.\ \emph{e}., on the same
side as in figs.6b,c, or on both sides as in fig.6d. On the opposite,
processes in which one ``out'' boson is made with the same fermions as
one of the ``in'' bosons can be constructed from the exchange diagram of
fig.2b, one of the two direct Coulomb scatterings having however to
``touch'' this unchanged pair, as in figs.6e,f, in order to have this
composite boson changing state.

\subsection{Some general comments based on dimensional arguments}

The qualitative analysis of what can possibly happen to two or three
composite bosons has led us to draw very many possible processes able
to make them changing states. It is however of importance to note that
all these complicated processes can be constructed just with two
elementary blocks, $\lambda_{mnij}$ and $\xi_{mnij}$, through $L_2
\left(_m^n\ _i^j\right)$ and $C\left(_m^n\ _i^j\right)$, \emph{i}.\ 
\emph{e}., a pure fermion exchange and a clean direct Coulomb
interaction
\emph{between} two composite bosons --- which is the only
process unambiguously between the composite bosons of
both sides

$\xi_{mnij}$ is a scattering in the usual sense, \emph{i}.\ \emph{e}.,
it has the dimension of an energy. This in particular means that each
time a new $\xi_{mnij}$ appears in a physical quantity, a new energy
denominator has also to appear; on the opposite, $\lambda_{mnij}$ is an
unconventional ``scattering'' because it is dimensionless. In addition,
depending on the way a new Pauli scattering appears, it can either
``kill'' the preceding one as in eq.(2.10), or help to mix more
composite bosons as in eq.(2.27).

With respect to the possible goals of a many-body expansion, this makes
them playing very different roles. If the relevant energies are the
detunings --- as in problems dealing with optical nonlinearities ---
the energy denominator which appears with a new $\xi_{mnij}$ is
made of detunings, so that, for unabsorbed photons, \emph{i}.\
\emph{e}., large detuning, we just have to look for processes in
which enters the smallest amount of
$\xi$'s.

If we are interested in density effects, this is more subtle. The
dominant terms at small density are dominated by processes in which
enters the smallest amount of particles, \emph{i}.\ \emph{e}.,
diagrams with the smallest amount of lines. While, in the case of
elementary bosons, we need one scattering to connect two lines, two
scatterings to connect three lines, and so on\ldots (see figs.1b,f), so
that each new line goes with a new energy denominator, this is no more
true for composite bosons: Indeed, we can connect lines in the absence
of any Coulomb scattering. Moreover, while, with exchanges alone, for
two lines we need one Pauli scattering and for three lines we need two,
these Pauli scatterings have to be put in very specific positions 
not to ``destroy''themselves. Consequently, in order to generate a
density expansion, in a system of composite bosons, to look at the
number of $\xi$ or $\lambda$ scatterings does not
really help. We should, instead, start with the appropriate number of
composite-boson lines (two for terms at lowest order in density, three
for the next order terms, and so on \ldots) and construct the possible
connections between these lines, using $\lambda_{mnij}$ and/or
$\xi_{mnij}$.

Of course, all this can be qualified of wishful thinking or handwaving
arguments. These qualitative remarks are however of great help to
identify the physics we want to describe through its visualization in
this new set of diagrams. A hard mathematical derivation of all these
intuitive thinkings can always be recovered by calculating the physical
quantity at hand, expressed in terms of composite boson operators,
through matrix elements like\linebreak
$\langle v|B_{m_N}\cdots B_{m_1}\,f(H)\,B_{i_1}^\dag\cdots
B_{i_N}^\dag|v\rangle$. To calculate them, the Hamiltonian depending
quantity
$f(H)$ is first pushed to the right using
$[f(H),B_i^\dag]$ which can be deduced from eqs.(2.15,17) for any type
of function $f$. This makes appearing a set of direct scatterings
$\xi_{mnij}$. The remaining scalar product of
$N$-composite-boson states is then calculated using eqs.(2.3,5). This
makes appearing a set of Pauli scatterings $\lambda_{mnij}$. Note
that, in this procedure, the $\xi$'s are all together on the right,
while the
$\lambda$'s are all together on the left (or the reverse if we push
$f(H)$ to the left). This in particular avoids spurious mixtures of
$\xi$'s and $\lambda$'s like the one of fig.3g.

\section{Conceptual problems with bosonization}

It is of course an appealing idea to try to find a way to replace
composite bosons by elementary bosons, because textbook techniques can
then be used to treat their many-body effects. In view of section 2, it
is however clear that such a replacement raises various problems:

(i) While elementary-boson states are orthogonal, the composite boson
ones are not (see eqs.(2.7,27)).

(ii) This is linked to the fact that, while elementary-boson states
form a complete set, the set of composite-boson states is overcomplete.

(iii) While only one elementary scattering between two elementary
bosons exists, namely $\xi_{mnij}^\mathrm{eff}$, in the case of
composite bosons, we have identified three scatterings with the
dimension of an energy, namely $\xi_{mnij}$, $\xi_{mnij}^\mathrm{in}$
and $\xi_{mnij}^\mathrm{out}$, plus one dimensionless scattering
$\lambda_{mnij}$, this last scattering in fact allowing to construct 
$\xi_{mnij}^\mathrm{in}$ and $\xi_{mnij}^\mathrm{out}$ from
$\xi_{mnij}$. Consequently, between composite bosons, there are
two fully independent scatterings --- elementary bosons having one only.

(iv) While all the complicated processes which can exist with three
composite bosons can be decomposed in terms of $\xi_{mnij}$ and
$\lambda_{mnij}$, it is necessary to introduce additional potentials
between three elementary bosons in the Hamiltonian, if we want to take
care of them. And so on, if we are interested in processes involving
four, five,\ldots bosons, \emph{i}.\ \emph{e}., in higher order terms in
the boson density.

Among all these problems, the overcompleteness of composite-boson
states is for sure the major one. Let us consider it at first.

\subsection{Nonorthogonality and overcompleteness}

These two problems are of course linked, the overcompleteness generating
the nonorthogonality of the composite-boson states. However, the
overcompleteness is far more difficult to handle. Just to grasp the
difficulty, consider a 2D plane. To represent it, we can use the standard
orthogonal basis $(\v x,\v y)$ but we can as well use any two vectors
$(\v x',\v y')$ which are not colinear. From them, we can either
construct two orthogonal vectors, for example $(\v x'',\v y')$, with 
$\v x''=\v x'-(\v x'.\v y')\v y'$, or we can just keep them. This will
make the algebra slightly more complicated because $\v x'.\v y'\neq 0$,
but that's all. If it now happens that three vectors of the 2D plane,
$(\v x',\v y',\v z')$ are equally relevant, so that there is no good
reason to eleminate one, then we must find a good way to mix them in
order to produce two vectors out of three, which can serve as a basis
for the 2D plane.

In the case of bosons, the space dimension is of course infinite, as
well as the number of ``unnecessary'' states, so that the space
reduction cannot be an easy task. On that respect, to face the
overcompleteness of the composite-boson states and to handle it, as we
do, up to the end, seems to us a very secure way to control all types of
tricky many-body effects between composite bosons.

If we just consider the problem of nonorthogonality, we can think of
overcoming it by considering a physically relevant $N$-composite-boson
state, for example $|0\rangle=B_0^{\dag N}|v\rangle$, with all the
bosons in the same state (this state is close to the
$N$-composite-boson ground state). We can then replace the other
composite-boson states, for example $|I\rangle=B_i^\dag B_0^{\dag N-1}
|v\rangle$, by their component perpendicular to $|0\rangle$, namely
$|I'\rangle=P_{\perp}B_i^\dag B_0^{\dag N-1}|v\rangle$, where
\begin{equation}
P_{\perp}=1-\frac{|0\rangle\langle 0|}{\langle 0|0\rangle}\ .
\end{equation}
This actually helps partly only, because, even if we now have $\langle
0|I'\rangle=0$, these $|I'\rangle$ states are not really good in the
sense that they do not form an orthogonal set: We still have
$\langle J'|I'\rangle\neq 0$. This remaining nonorthogonality can be
unimportant in problems in which the $\langle J'|I'\rangle$ scalar
products do not appear --- as in cases in which they correspond to
``higher order terms''. However, even in these cases, such a
construction of an orthogonal set is clearly not fully satisfactory,
when compared to handling the nonorthogonality, really.

\subsection{``Good'' effective scattering}

Our study of the interactions between two composite bosons makes
appearing four scatterings: $\xi_{mnij}$, $\xi_{mnij}^\mathrm{in}$,
$\xi_{mnij}^\mathrm{out}$ and $\lambda_{mnij}$. Let us, for a while,
accept the idea to have bosonized particles which form an orthogonal
set, so that the pure Pauli scatterings do not play a role, \emph{i}.\
\emph{e}., we drop all the $\lambda_{mnij}$'s. We are left with three
scatterings having the dimension of an energy. An idea for a ``good''
effective scattering between elementary bosons can be to have the same
Hamiltonian matrix elements within the two-boson subspace. However, in
view of eqs. (1.8) and (2.19), we are in trouble if we keep dropping
the $\lambda_{mnij}$'s, because we can choose either
$\xi_{mnij}-\xi_{mnij}^\mathrm{in}$ or
$\xi_{mnij}-\xi_{mnij}^\mathrm{out}$, these two quantities being equal
for $E_m+E_n=E_i+E_j$ only, due to eq.(2.21). If, instead, we keep the
$\lambda_{mnij}$'s, we are led to take
\begin{equation}
\hat{\xi}_{mnij}^\mathrm{eff}=\xi_{mnij}-\left[\xi_{mnij}^\mathrm{in}+
(E_i+E_j)\lambda_{mnij}\right]\ ,
\end{equation}
with the bracket possibly replaced by $\left[\xi_{mnij}^\mathrm{out}
+(E_m+E_n)\lambda_{mnij}\right]$; so that we can rewrite this effective
scattering, in a more symmetrical form, as
\begin{equation}
\hat{\xi}_{mnij}^\mathrm{eff}=\xi_{mnij}-\frac{1}{2}\left[\xi_{mnij}
^\mathrm{in}
+\xi_{mnij}^\mathrm{out}+(E_m+E_n-E_i-E_j)\lambda_{mnij}\right]\ .
\end{equation}
We note that this $\hat{\xi}_{mnij}^\mathrm{eff}$ is such that
$\hat{\xi}_{mnij}^\mathrm{eff}=\left(\hat{\xi}_{ijmn}^\mathrm{eff}
\right)^\ast$,
as necessary for the hermiticity of the effective Hamiltonian for
elementary bosons. If we now decide to drop the Pauli
scatterings $\lambda_{mnij}$'s, we are led to take
\begin{equation}
\xi_{mnij}^\mathrm{eff}=\xi_{mnij}-(\xi_{mnij}^\mathrm{in}
+\xi_{mnij}^\mathrm{out})/2\ ,
\end{equation}
which preserves the hermiticity of the Hamiltonian. This has to be
contrasted with the effective scattering for bosonized excitons
extensively used by the semiconductor community [11,12], namely
$\xi_{mnij}-
\xi_{mnij}^\mathrm{out}$, as first obtained by Hanamura and Haug,
following an Inui's bosonization procedure [20].

Before going further, let us note that, in dropping the $\lambda_{mnij}$
term in $\hat{\xi}^\mathrm{eff}$ to get $\xi^\mathrm{eff}$, we actually
``drop'' a quite unpleasant feature of this effective scattering: its
spurious dependence on the band gap in the case of excitons. Indeed, in
$\hat{\xi}_{mnij}^\mathrm{eff}$ appears the sum --- not the
difference --- of the ``in'' and ``out'' boson energies. In the case of
excitons, this boson energy is essentially equal to the band gap plus a
small term depending of the particular exciton state considered. So that
$E_m+E_n+E_i+E_j$ is essentially equal to four times the band gap. Its
appearance in a scattering is a physical nonsense.

All this leads us to conclude that the only ``reasonable'' scattering
between two elementary bosons --- which has the dimension of an
energy, preserves hermiticity and has no spurious band gap dependence
--- should be $\xi_{mnij}^\mathrm{eff}$.

Actually, even this $\xi_{mnij}^\mathrm{eff}$ is not good, except
may be for effects in which only enter \emph{first} order
\emph{diagonal} Coulomb processes --- in order for the ``in'' and
``out'' Coulomb scatterings to be equal. Indeed, in a previous work
[10], we have shown that the link between the inverse lifetime of an
exciton state --- due to exciton-exciton interations --- and the sum of
its scattering rates towards a different exciton state, misses a factor
of 2, if the excitons are replaced by elementary bosons, \emph{whatever}
is the effective scattering used --- a quite strong statement! We have
recently recovered this result [21], without calculating the two
quantities explicitly, just by using an argument based on differences
in the closure relations of elementary and composite excitons.

Let us now come back to the problem of having the Pauli scatterings
systematically missing in any approach which uses an effective
Hamiltonian. It is actually far worse than the problem of a ``good''
exchange part for Coulomb scattering, because we not only miss a factor
of 2, but the dominant term [15,16] in all optical nonlinear effects!
Indeed, a photon interacts with a semiconductor through the virtual
exciton to which this photon is coupled. If the semiconductor already has
excitons, the first way this virtual exciton interacts is via Pauli
exclusion, since this exclusion among fermions makes it filling all the
fermion states already occupied in the sample. Coulomb interaction
comes next, since it has to come with an energy denominator which, in
problems involving photons, is a detuning, so that these Coulomb terms
always give a negligible contribution at large detuning, in front of the
terms coming from Pauli scatterings alone.

Beside the exciton optical Stark effect, in which the roots of our
many-body theory for composite excitons can be found [22], we have
studied some other optical nonlinearities in which the interaction of a
composite exciton with the matter is dominated by Pauli scattering,
namely the theory of the third order nonlinear susceptibility
$\chi^{(3)}$ [16], the theory of Faraday rotation [23] and the
precession of a spin pined on an impurity [24].

Since this Pauli scattering, quite crucial in many physical effects, is
dimensionless, it cannot appear in the effective Hamiltonian of
bosonized particles, which needs a scattering having the dimension of
an energy. Consequently, all terms in which this scattering appears
alone, \emph{i}.\ \emph{e}., not mixed with Coulomb, are going to be
missed in any procedure using an effective Hamiltonian. (This is also
true for approaches using spin-spin Hamiltonians [13]).

Finally, our qualitative discussion on the possible interactions
between three composite bosons, has led us to identify, in addition to
pure exchange processes based on $L_3$, again missed, more complicated
mixtures of Coulomb and exchange than the one appearing between two
composite bosons, $\xi_{mnij}^\mathrm{in}$ and $\xi_{mnij}
^\mathrm{out}$. In order not to miss them, we could think of adding a
three-body part to the Hamiltonian like
\begin{equation}
\bar{V}^{(3)}=\frac{1}{3!}\sum_{mnpijk}\bar{\xi}_{mnpijk}^\mathrm{eff}\,
\bar{B}_m^\dag\bar{B}_n^\dag\bar{B}_p^\dag\bar{B}_i\bar{B}_j\bar{B}_k
\ . 
\end{equation}
Let us however note that the proper identification of
$\bar{\xi}_{mnpijk} ^\mathrm{eff}$ with the three-body processes
which cannot be constructed from $\xi_{mnij}$,
$\xi_{mnij}^\mathrm{in}$ and $\xi_{mnij} ^\mathrm{out}$, is not fully
straightforward because this three-body potential $\bar{V}^{(3)}$
formally contains terms in which one elementary boson can stay
unchanged, \emph{i}.\ \emph{e}., terms already included in
$\bar{V}^{(2)}$.

All this actually means that the ``good'' effective Hamiltonian, apart
from the pure Pauli terms which are going to be missed anyway, has to
be more and more complicated if we want to include processes in which
more and more bosons are involved, \emph{i}.\ \emph{e}., if we want to
study many-body effects, really. Just for that, the replacement of
composite bosons by elementary boson seems to us far more complicated
than keeping the boson composite nature through a set of Pauli
scatterings, as we propose.

\section{Extension to more complicated composite bosons}

In the preceding sections, we have considered composite bosons made of
a pair of different fermions, these pairs being eigenstates of the
Hamiltonian. In this last section, we are going to show how we can
generalize the definitions of the various scatterings we have found, to
the case of pairs of fermions which are not Hamiltonian eigenstates.
For clarity, we are going to show this generalization on a specific
example of current interest: a composite boson made of a pair of
trapped electrons [13,14].

Let us consider two electrons with two traps located at $\v R_1$ and
$\v R_2$. These traps can be semiconductor quantum dots, Coulomb traps
such as ionized impurities, H atom protons, and so on\ldots The system
Hamiltonian then reads
\begin{equation}
H=H_0+V_{ee}+W_{\v R_1}+W_{\v R_2}\ ,
\end{equation}
where $H_0$ is the kinetic contribution, $V_{ee}$ the electron-electron
Coulomb interaction and $W_{\v R}$ the potential of the trap located at
$\v R$. The physically relevant one-electron states [24] are the
one-electron eigenstates in the presence of one trap located at $\v R$,
namely $|\v R\mu\rangle$ given by $(H_0+W_{\v R}-\epsilon_\mu)|\v R\mu
\rangle=0$. They are such that
\begin{equation}
|\v R\mu\rangle=a_{\v R\mu}^\dag|v\rangle=\sum_{\v k}\langle\v k|\v R\mu
\rangle\,a_{\v k}^\dag|v\rangle\ ,
\end{equation}
$a_{\v k}^\dag$ being the creation operator for a free electron with
momentum $\v k$. In the case of Coulomb trap, the $|\v R\mu\rangle$
states are just the H atom bound and extended states.

We now consider the two-electron states having one electron on each
trap,
\begin{equation}
|n\rangle=A_n^\dag|v\rangle=a_{\v R_1\mu_1}^\dag\,a_{\v R_2\mu_2}^\dag
|v\rangle\ .
\end{equation}
These states do not form an orthogonal set since, due to the finite
overlap of the one-electron wave functions, we do have
\begin{equation}
\langle n'|n\rangle=\delta_{n',n}-\lambda_{n'n}^{(e-e)}\ ,
\end{equation}
where $\lambda_{n'n}^{(e-e)}=\langle\v R_1\mu'_1|\v R_2\mu_2\rangle\,
\langle\v R_2\mu'_2|\v R_1\mu_1\rangle$. This possible carrier exchange
between the two traps, shown in fig.7a, produces not only the
nonorthogonality of the $|n\rangle$ states, but also the
overcompleteness of this set of states. Indeed, by putting the electron
of the $\v R_1$ trap in a state of the $\v R_2$ trap, we can show that
\begin{equation}
A_n^\dag=-\sum_{n'}\lambda_{n'n}^{(e-e)}\,A_{n'}^\dag\ .
\end{equation}

If we now want to determine the Pauli scatterings of this composite
boson made of a pair of trapped electrons, we are led to define the
deviation-from-boson operator $D_{n'n}$ through
\begin{equation}
D_{n'n}=\langle n'|n\rangle-[A_{n'},A_n^\dag]\ ,
\end{equation}
which is a generalization of eq.(2.3) to the case of nonorthogonal
composite bosons. Indeed, with such a definition, we still have the now
standard property of a deviation-from-boson operator, namely $D_{n'n}
|v\rangle=0$. The Pauli scatterings of the composite boson $A_n^\dag$
with another composite boson $B_i^\dag$ is then obtained through
\begin{equation}
[D_{n'n},B_i^\dag]=2\sum_{i'}\lambda_{n'i'ni}^{(ee-X)}\,B_{i'}^\dag\ .
\end{equation}
In a case of current interest, namely the spin manipulation by a laser
pulse [13,14,25,26], the relevant bosons $B_i^\dag$ with which the pair
of trapped electrons interact are the virtual excitons coupled to the
photons. This composite boson $B_i^\dag$, made of an electron-hole pair
can exchange its electron with one of the two electrons of the composite
boson $A_n^\dag$, through the Pauli scattering
$\lambda_{n'i'ni}^{(ee-X)}$ shown in fig.7b. This is why we have
defined it with a 2 prefactor in eq.(4.7).

We now look for a scattering of the composite boson $A_n^\dag$ having
the dimension of an energy. For that, we first note that $H_0+W_{\v R}$
can be written in terms of $a_{\v R\mu}^\dag$ as [24]
\begin{equation}
H_0+W_{\v R}=\sum_\mu\epsilon_\mu\,a_{\v R\mu}^\dag a_{\v R\mu}\ ,
\end{equation}
where $\epsilon_\mu$ is the energy of the one-electron state $|\v R\mu
\rangle$. This leads to
\begin{equation}
H|n\rangle=E_n|n\rangle+\sum_{n'}\xi_{n'n}^{(e-e)}|n'\rangle\ ,
\end{equation}
where $E_n=\epsilon_{\mu_1}+\epsilon_{\mu_2}$ is the ``free'' energy of
the pair of trapped electrons, while $\xi_{n'n}^{(e-e)}$ comes from
their Coulomb repulsion as well as to the interaction of each electron
with the other trap (see fig.7c).

For such a composite boson $A_n^\dag$, which, due to eq.(4.9), is not
eigenstate of the Hamiltonian, the proper way to define the ``creation
potential'' is through
\begin{equation}
V_n^\dag=[H,A_n^\dag]-E_nA_n^\dag-\sum_{n'}\xi_{n'n}^{(e-e)}A_{n'}^\dag
\ ,
\end{equation}
in order to still have $V_n^\dag|v\rangle=0$, eq.(4.10) being a
generalization of eq.(2.15). We then get the ``direct Coulomb
scattering'' between this composite boson $A_n^\dag$ and another
composite boson $B_i^\dag$, through
\begin{equation}
[V_n^\dag,B_i^\dag]=\sum_{n'i'}\xi_{n'i'ni}^{(ee-X)}\,A_{n'}^\dag B_{i'}
^\dag\ ,
\end{equation}
which is similar to eq.(2.17). This direct scattering is shown in
fig.7d. It corresponds to the direct Coulomb interaction of each of the
two trapped electrons with the electron-hole pair of the exciton.

Using this set of commutators and the two
scatterings $\lambda_{n'n}^{(e-e)}$ and
$\xi_{n'n}^{(e-e)}$, it is actually possible to calculate the energy of
two trapped electrons with their possible exchanges included exactly,
in order to determine the singlet-triplet splitting these exchange
processes induce in the van der Waals energy [27]. Using them and the
two scatterings
$\lambda_{n'n}^{(ee-X)}$ and $\xi_{n'n}^{(ee-X)}$, we can also calculate
the increase of this splitting induced by virtual excitons
coupled to a laser beam [28], resulting from additional electron
exchanges with the electron of the virtual exciton. This problem is of
great interest for the possible control of the spin transfer time
between two traps by a laser pulse, having in mind its possible use for
quantum information [29].

\section{Conclusion}

In this paper, we have made a detailed qualitative analysis of what can
be called ``interaction'' between two or three composite bosons. We have
shown that all the processes identified to produce a change in the
boson states can be written in terms of two blocks only: a direct 
Coulomb scattering which has the dimension of an energy and a pure
Pauli ``scattering'' which is dimensionless. This Pauli scattering is
actually the novel ingredient of our many-body theory for composite
bosons, in which these composite bosons are never replaced by
elementary bosons.

We can possibly think of including processes in which enter
complicated combinations of direct Coulomb scatterings and Pauli
scatterings through a set of effective scatterings between two, three,
or more elementary bosons. On the opposite, all processes in which the
Pauli scatterings appear alone have to be missed if one uses
effective Hamiltonians such as the ones in which the composite bosons
are replaced by elementary bosons, or any spin-spin Hamiltonian,
whatever the effective scattering is. This in particular happens in
all semiconductor optical nonlinearities, the virtual exciton coupled to
the photon field feeling the presence of the fermions present in the
sample, ``even more'' than their charges.

Finally, we have shown how to extend the mathematical definitions of
the Pauli scattering and the direct Pauli scattering to non trivial
composite bosons such as a pair of trapped electrons. This extension
again goes through the introduction of ``deviation-from-boson''
operators and ``creation potentials'', their main characteristics 
being to give zero when they act on vacuum, so that they really
describe interactions with the rest of the system.

Although it is easy to understand the reluctance one may have to enter
a new way of thinking interactions between composite bosons, we really
think that it is worthwhile to spend the necessary amount of time to
grasp these new ideas, in view of their potentiality in very many
problems of physics.

\newpage

\hbox to \hsize {\hfill FIGURE CAPTIONS
\hfill}

\noindent\textbf{Figure 1}

Basic diagrams for the interactions of two (a) and three (e) elementary
bosons.

Between two composite bosons, one (b), two (c), or more interations can
exist, while two (f) interactions at least are necessary to find three
composite bosons in ``out'' states (m,n,p) different from the ``in''
states (i,j,k).

Due to the boson undistinguishability, the elementary scattering
between two bosons must be invariant under a (m$\leftrightarrow$n)
and/or a (i$\leftrightarrow$j) permutation, as shown in (d).

\noindent\textbf{Figure 2}

(a) Basic diagram for the interaction of two composite bosons.

(b) Elementary hole exchange between these composite bosons.

(c) Elementary electron exchange between the same composite bosons as
the ones of (b). As shown in (c'), this process is equivalent to a hole
exchange with (m,n) changed into (n,m).

(d) Due to the undistinguishability of the fermions forming the
composite bosons, the elementary Pauli scattering $\lambda_{mnij}$
between two composite bosons must be invariant under a
(m$\leftrightarrow$n) and/or (i$\leftrightarrow$j) permutation. Due to
(c,c'), this Pauli scattering thus includes a hole exchange and an
electron exchange.

(e) Two hole exchanges reduce to an identity.

(f) One hole exchange followed by an electron exchange reduces to a
(m,n) permutation.

Note that all these processes are missed when composite bosons are
replaced by elementary bosons.

\noindent\textbf{Figure 3}

(a)Elementary \emph{direct} Coulomb interaction between two composite
bosons.

(b) This Coulomb interaction is made of e-e, h-h and two e-h
interactions.

(c) Due to the undistinguishability of the fermions forming the
composite bosons, the elementary direct Coulomb scattering $\xi_{mnij}$
between two composite bosons must be invariant under a
$(m\leftrightarrow n)$ and/or $(i\leftrightarrow j)$ permutation.

(d) The ``in'' Coulomb scattering $\xi_{mnij}^\mathrm{in}$ corresponds
to a direct Coulomb scattering followed by a carrier exchange. As shown
in (d'), the electron-hole Coulomb interaction of
$\xi_{mnij}^\mathrm{in}$ is
\emph{between} the ``in'' composite bosons, but inside the ``out'' ones.

(e) The ``out'' Coulomb scattering $\xi_{mnij}^\mathrm{out}$
corresponds to a carrier exchange followed by a direct Coulomb
interaction.

(f) Processes in which the direct Coulomb interaction is followed by
two hole exchanges reduce to a direct process.

(g) Processes in which the hole exchanges are on both
sides of the Coulomb direct interaction are physically strange because
their electron-hole parts
are ``inside'' both, the ``in'' and the ``out'' composite bosons, so
that they are already counted in these composite bosons: We never
find these strange processes appearing in physical effects resulting
from interactions between composite bosons.

\noindent\textbf{Figure 4}

(a) Basic diagram for the interaction of three composite bosons.

(b) ``Skeleton diagram'' for carrier exchange between three composite
bosons. It can be redrawn as in figs.(c,d): In all these diagrams, the
m composite boson has the same electron as i and the same hole as j.

(e) The skeleton diagram with the electron-hole lines exchanged
corresponds to a permutation of the boson indices.

The skeleton diagram between three composite bosons (b) can be drawn as
a succession of carrier exchanges between two composite bosons. Indeed,
(c) is nothing but (f), while (d) is nothing but (g).

(h) Due to the undistinguishability of the fermions forming the
composite bosons, the elementary Pauli scattering $\lambda_{mnpijk}$
between three composite bosons must be invariant under a $(m,n,p)$
and/or
$(i,j,k)$ permutation. It thus contains the $6\times 2=12$ processes
shown on this figure.

Note that, in the case of elementary bosons, two Coulomb interactions
at least are necessary to have all three bosons changing state, so that
these pure exchange terms are systematically missed when composite
bosons are replaced by elementary bosons.

\noindent\textbf{Figure 5}

Processes in which enter \emph{one} direct Coulomb scattering. In order
to have all three composite bosons changing state, these processes must
also contain one (a) or two (b) carrier exchanges. Note that such
processes with \emph{one} Coulomb interaction only do not exist for
elementary bosons, so that they are systematically missed when
composite bosons are replaced by elementary bosons.

\noindent\textbf{Figure 6}

Processes in which enter \emph{two} Coulomb interactions, either through
direct scatterings as in (a), or through a mixture of direct and
exchange processes as in (b-f). All these processes can be written in
terms of the two basic scatterings for composite bosons, namely the
direct Coulomb scattering $\xi_{mnij}$ and the Pauli scattering
$\lambda_{mnij}$.

\noindent\textbf{Figure 7}

(a)Pauli scattering $\lambda_{n'n}^\mathrm{(e-e)}$ between electrons
trapped in $\v R_1$ and $\v R_2$. In this exchange, the electrons can
end in trapped states $n'=(\mu'_1,\mu'_2)$ different from the initial
ones $n=(\mu_1,\mu_2)$.

(b) Pauli scattering $\lambda_{n'i'ni}^\mathrm{(ee-X)}$ between a
composite boson made of a trapped electron pair and a composite boson
made of an electron-hole pair, \emph{i}.\ {e}., more precisely a
composite exciton, their states changing from $(n,i)$ to $(n',i')$.

(c) Direct scattering $\xi_{n'n}^\mathrm{(e-e)}$ between two trapped
electrons. This scattering contains the Coulomb interaction between the
two electrons as well as the interactions of each electron with the
potential of the other trap.

(d) Direct scattering $\xi_{n'i'ni}^\mathrm{(ee-X)}$ between a
composite boson made of a trapped electron pair and a composite
exciton. This scattering contains the Coulomb interaction of the
exciton with each of the two trapped electrons.

\newpage

\begin{figure}
\centerline{ \scalebox{0.7}{\includegraphics{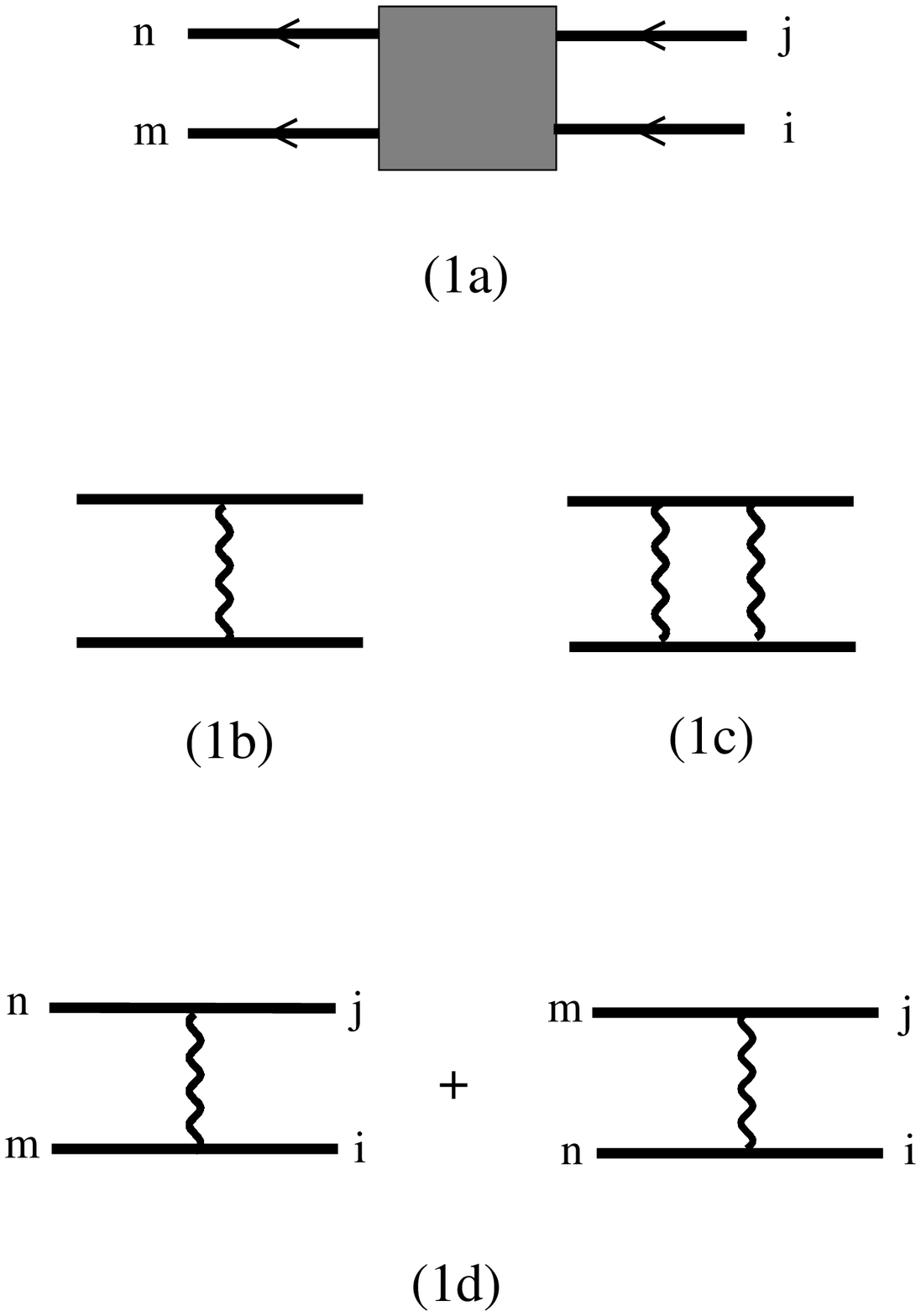}}}
\end{figure}

\clearpage

\begin{figure}
\centerline{ \scalebox{0.7}{\includegraphics{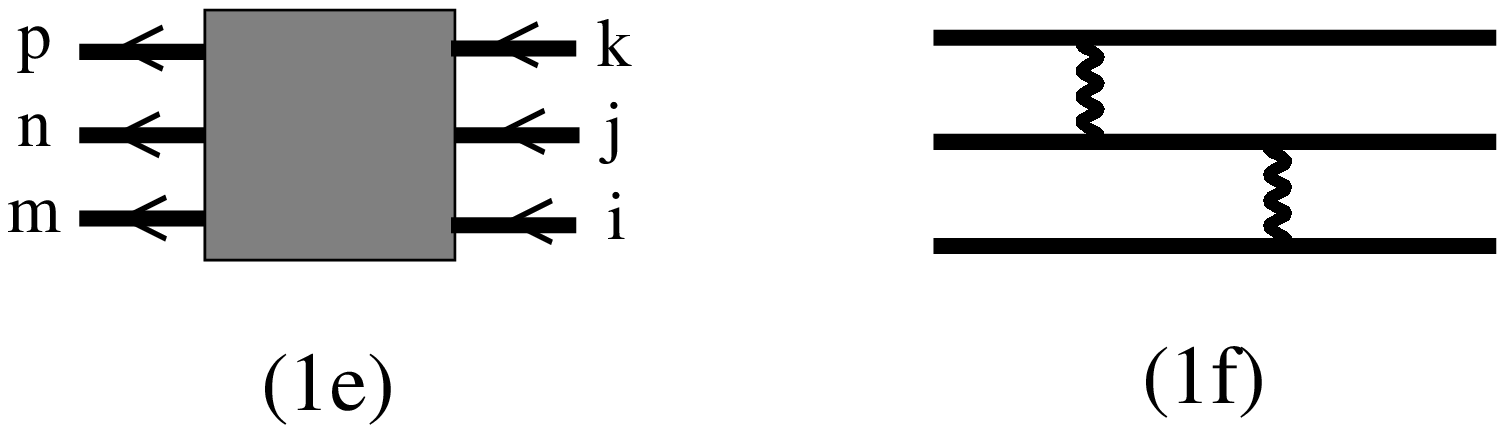}}}
\end{figure}

\clearpage

\begin{figure}
\centerline{ \scalebox{0.7}{\includegraphics{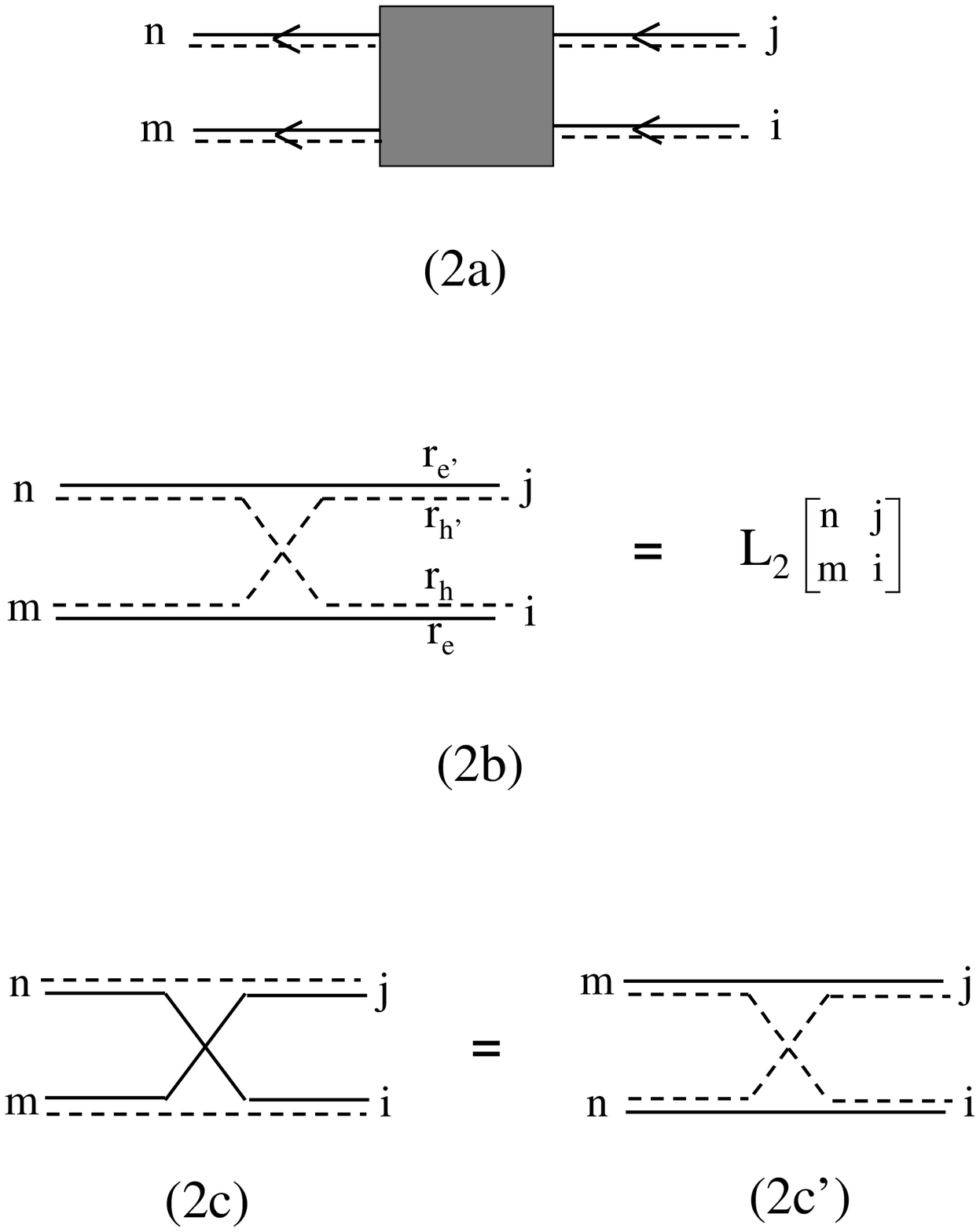}}}
\end{figure}

\clearpage

\begin{figure}
\centerline{ \scalebox{0.7}{\includegraphics{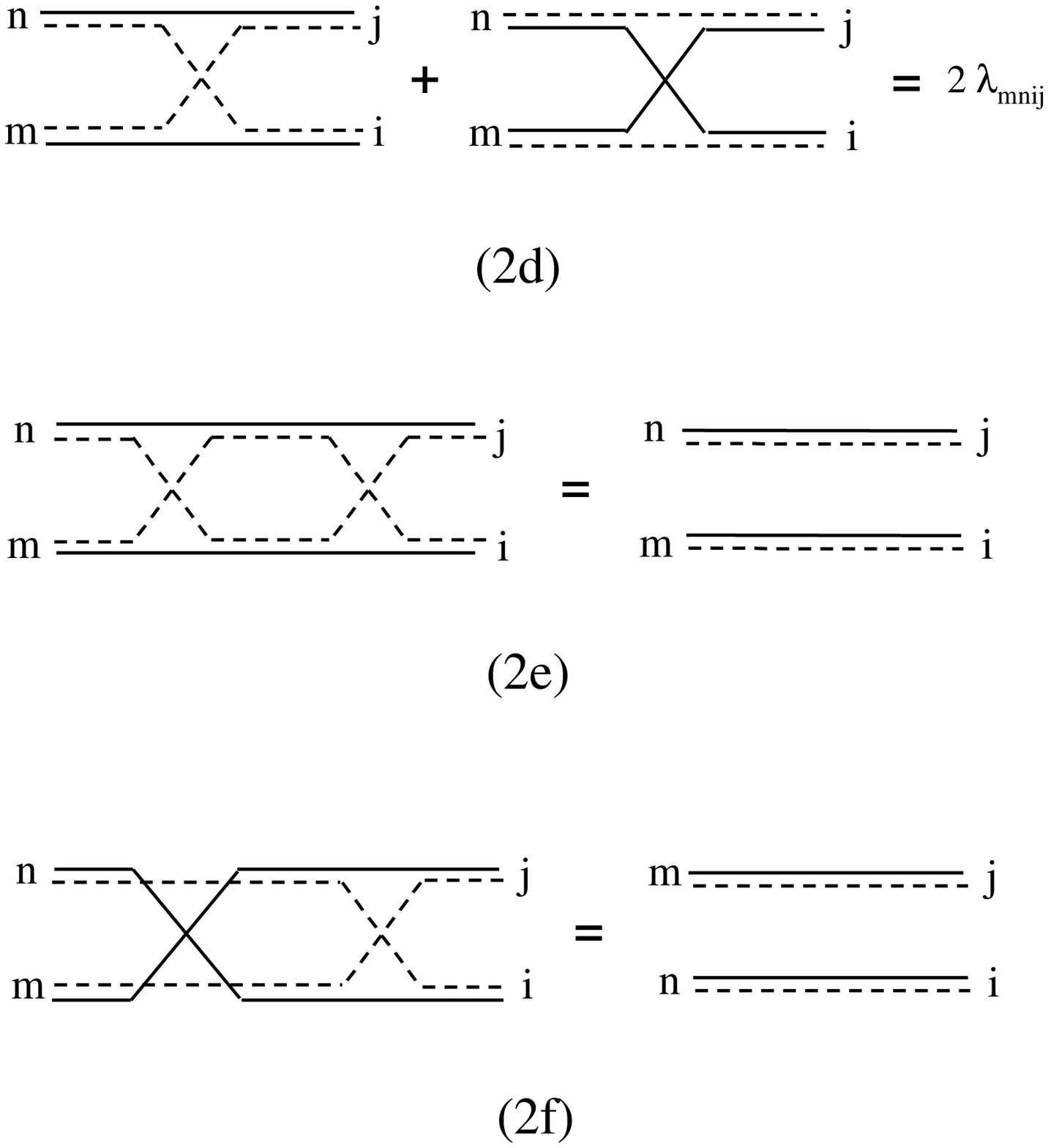}}}
\end{figure}

\clearpage

\begin{figure}
\centerline{ \scalebox{0.7}{\includegraphics{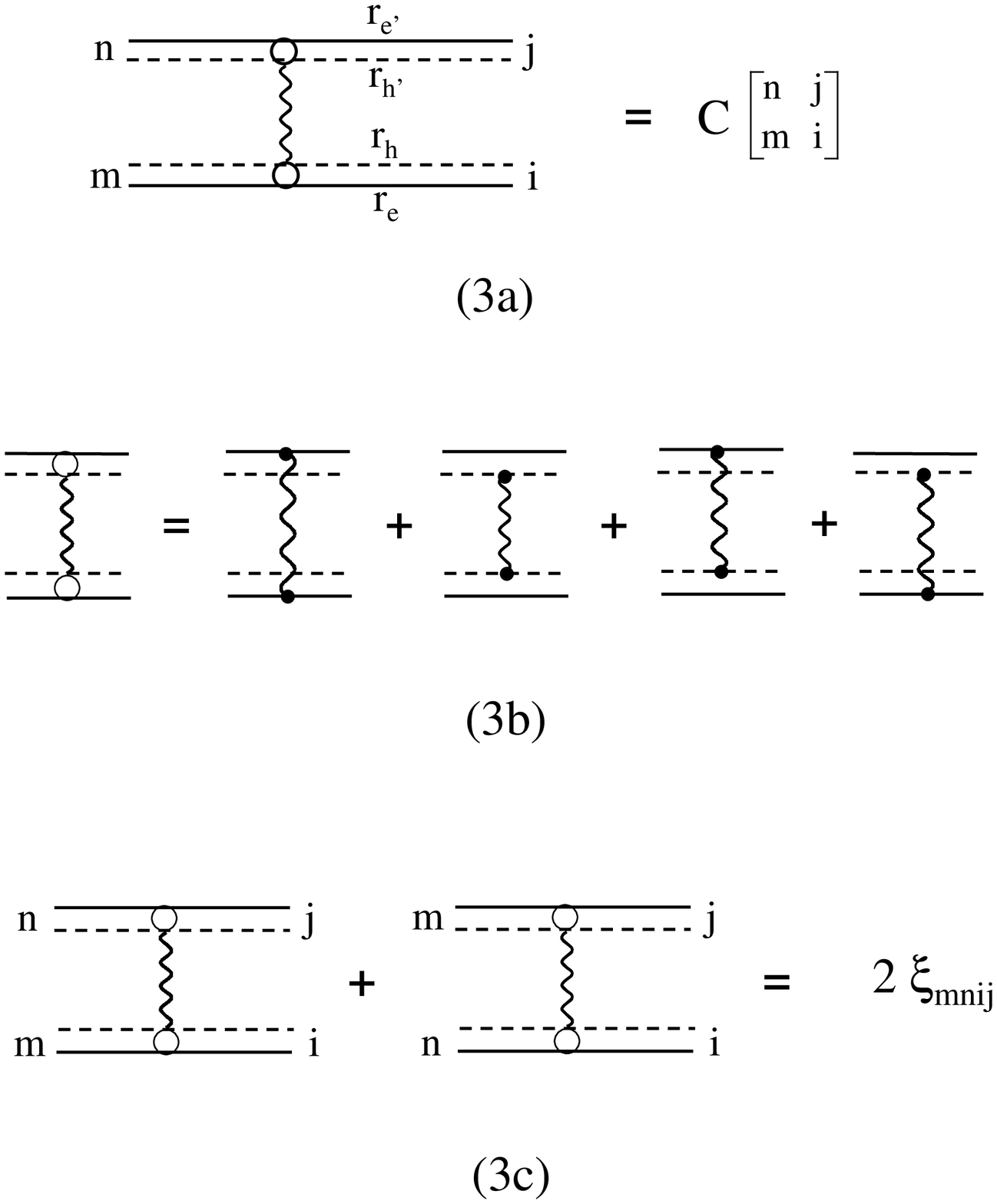}}}
\end{figure}

\clearpage

\begin{figure}
\centerline{ \scalebox{0.7}{\includegraphics{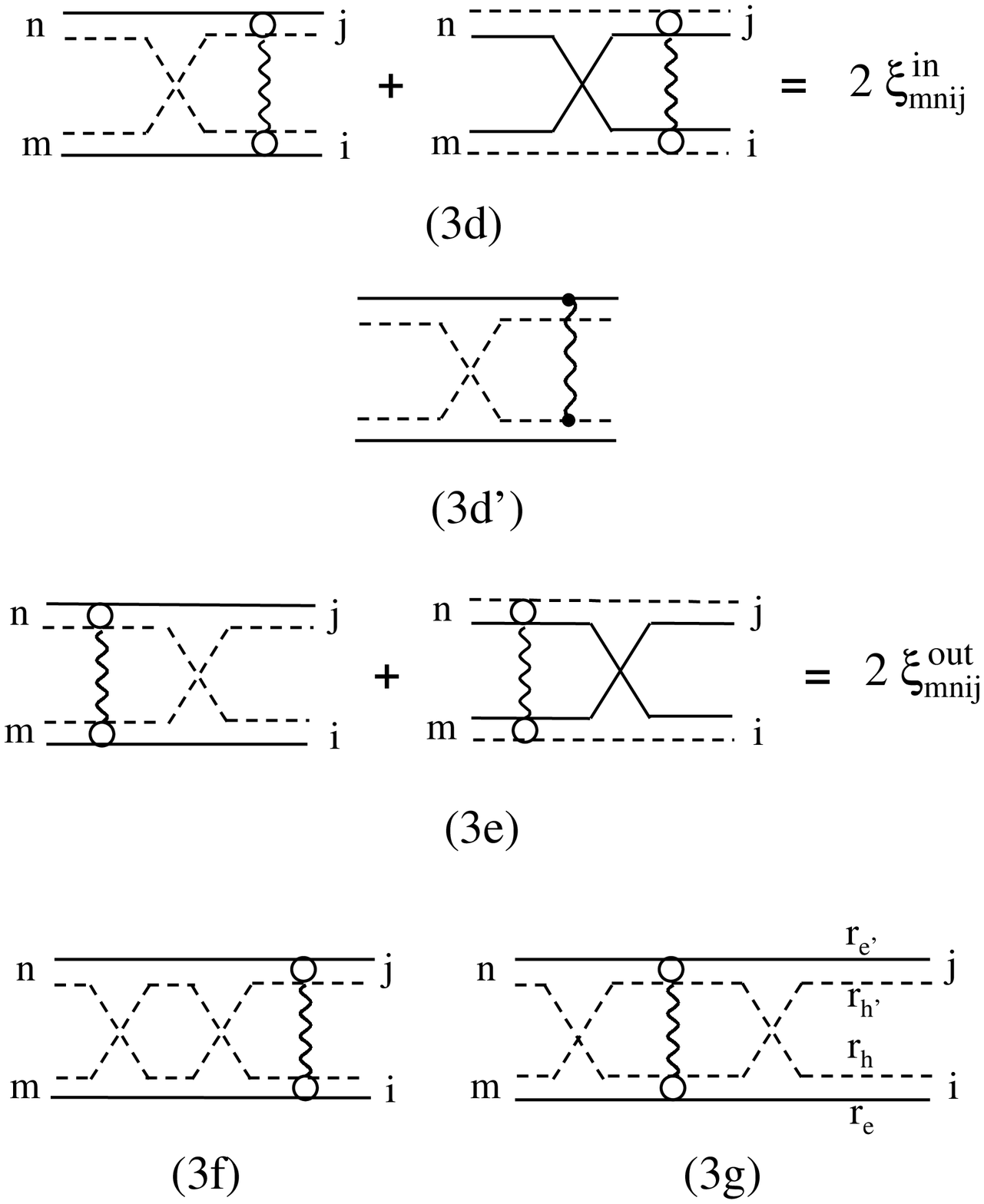}}}
\end{figure}

\clearpage

\begin{figure}
\centerline{ \scalebox{0.7}{\includegraphics{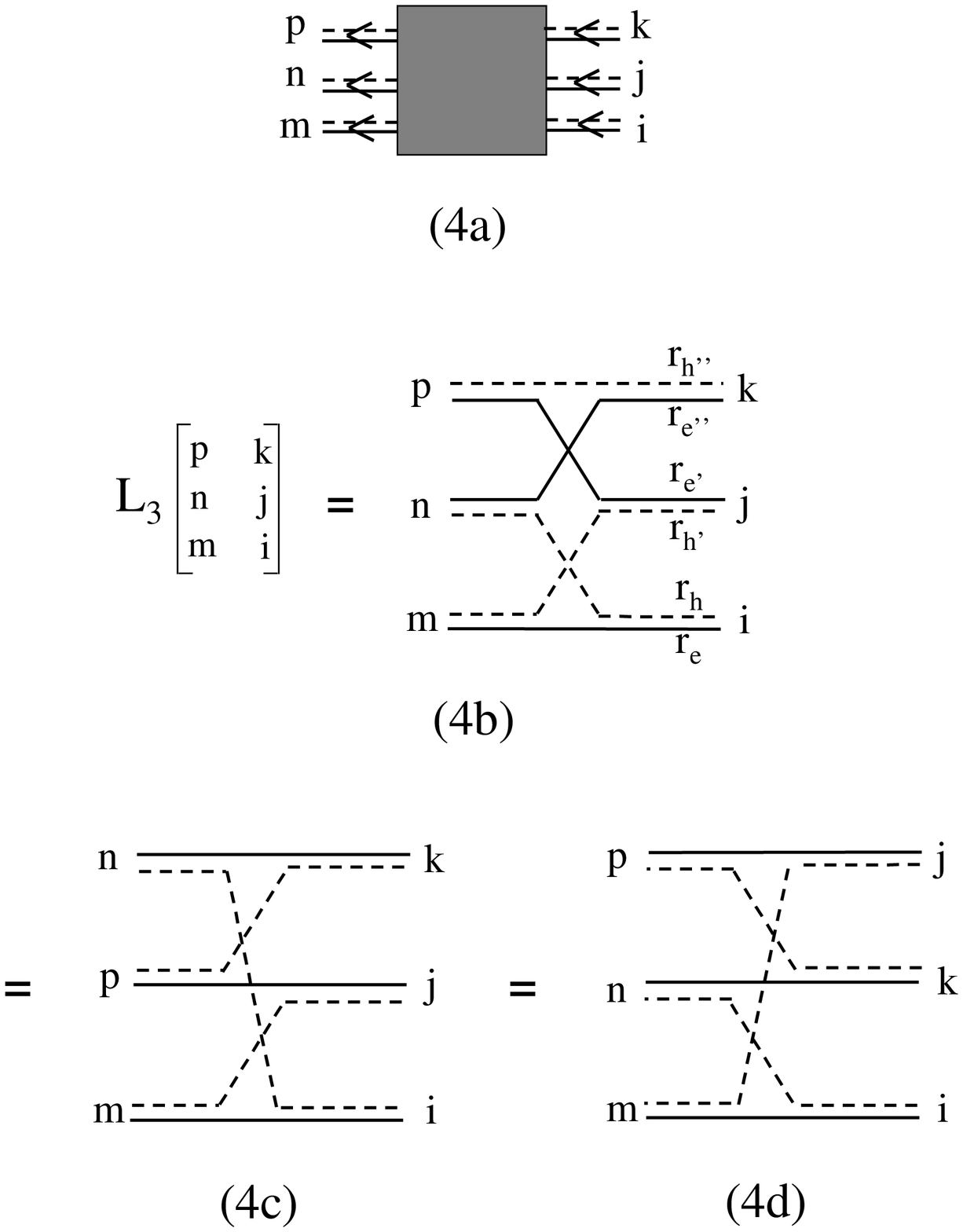}}}
\end{figure}

\clearpage
 
\begin{figure}
\centerline{ \scalebox{0.7}{\includegraphics{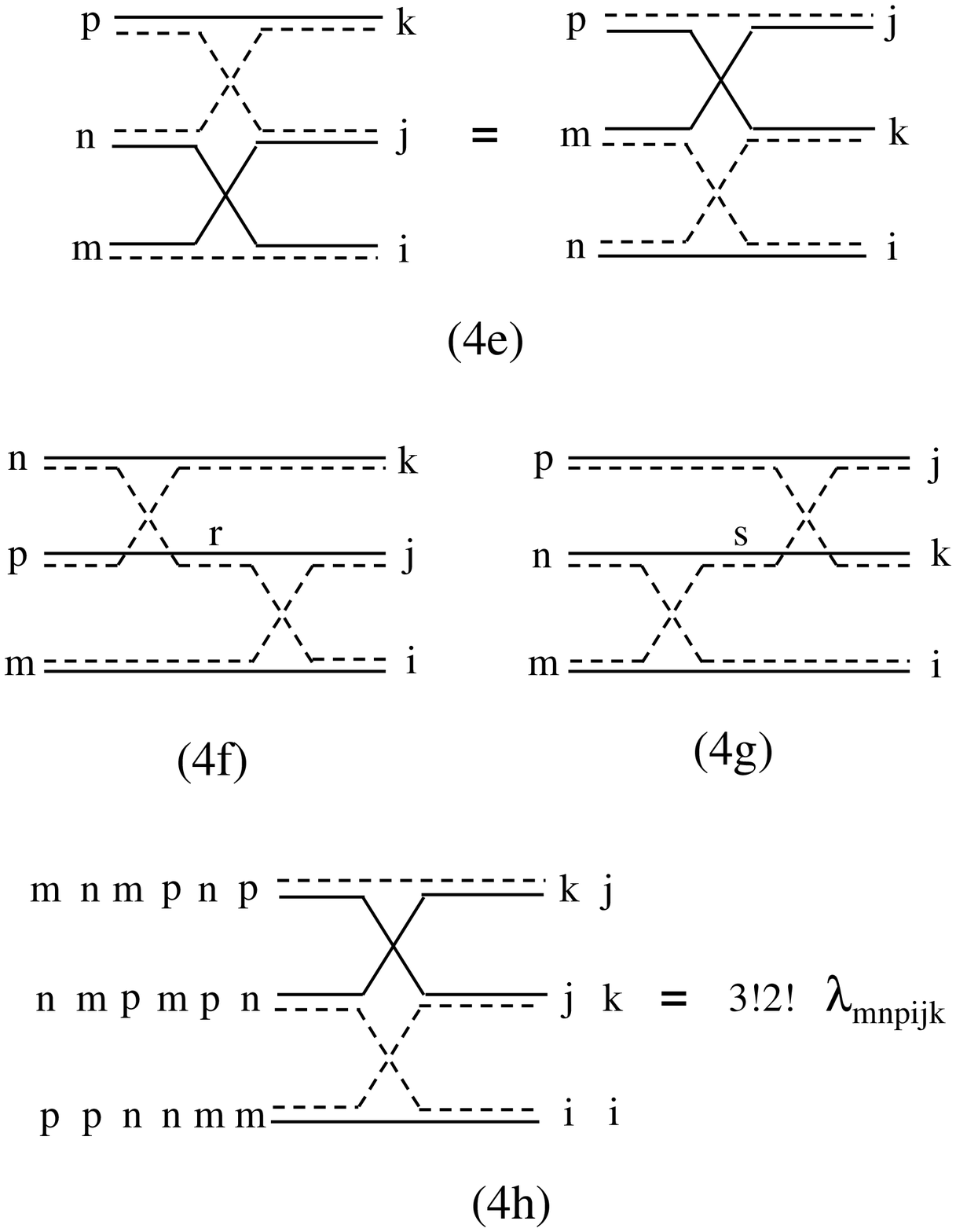}}}
\end{figure}

\clearpage

\begin{figure}
\centerline{ \scalebox{0.7}{\includegraphics{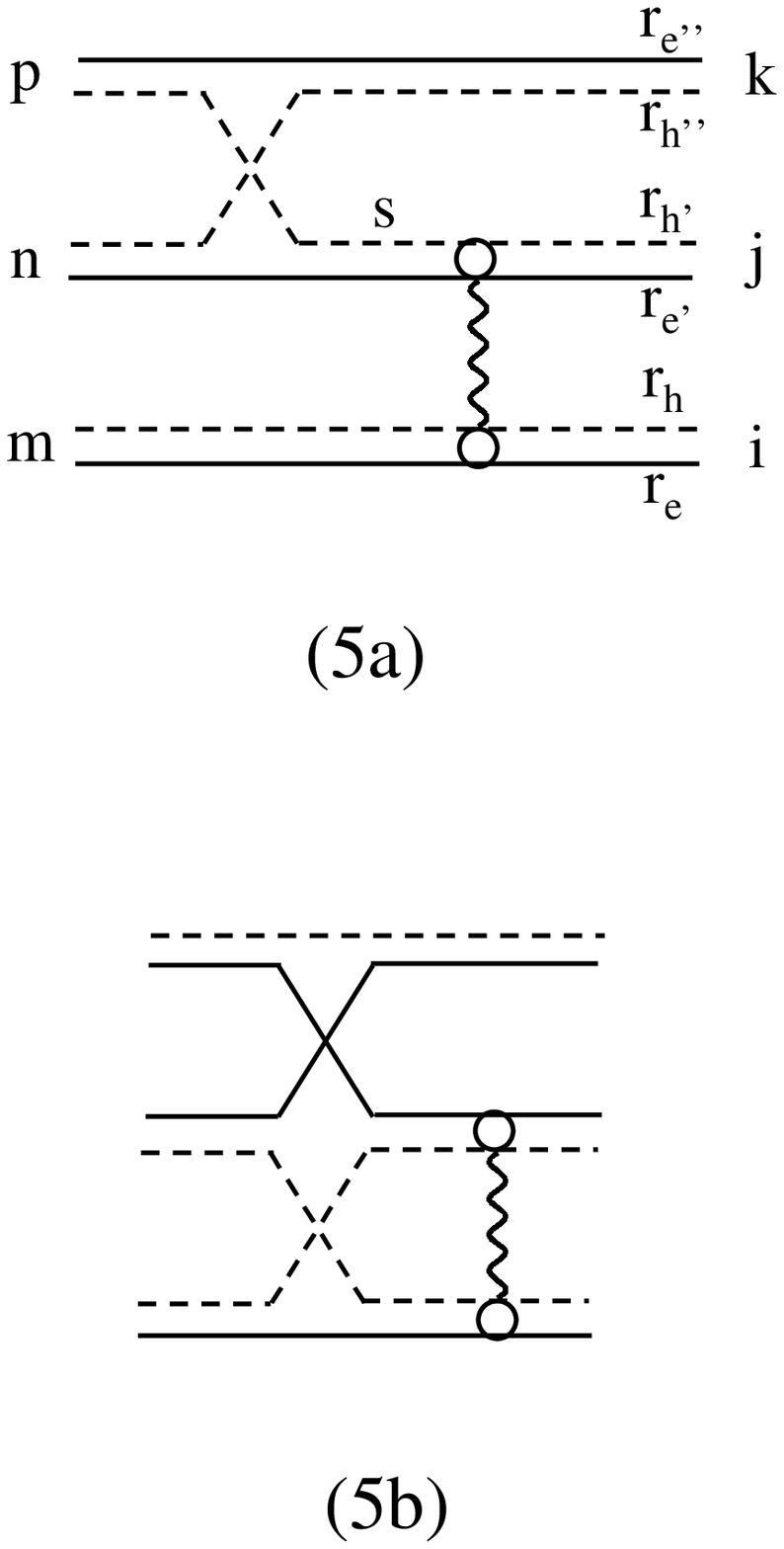}}}
\end{figure}

\clearpage

\begin{figure}
\centerline{ \scalebox{0.7}{\includegraphics{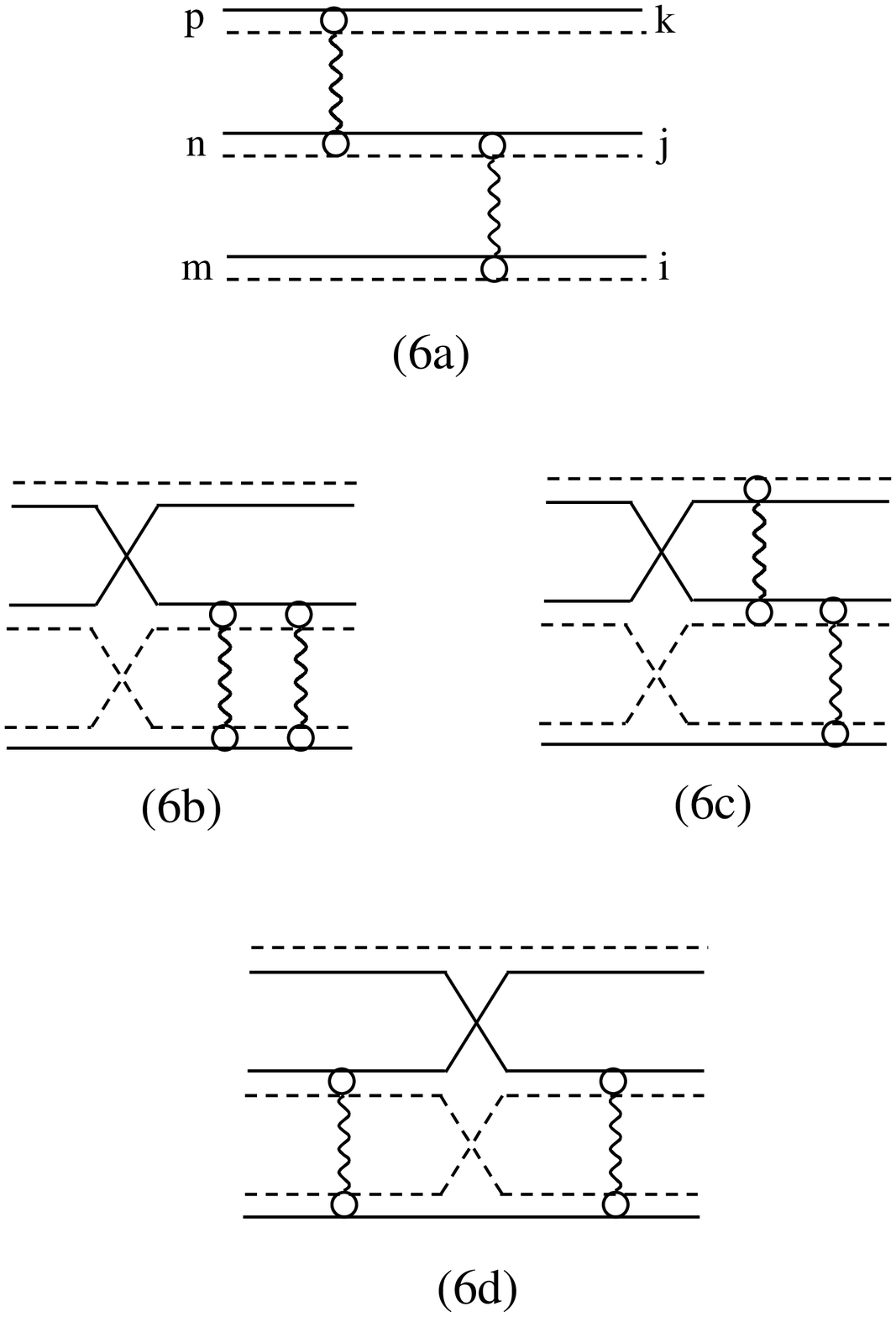}}}
\end{figure}

\clearpage

\begin{figure}
\centerline{ \scalebox{0.7}{\includegraphics{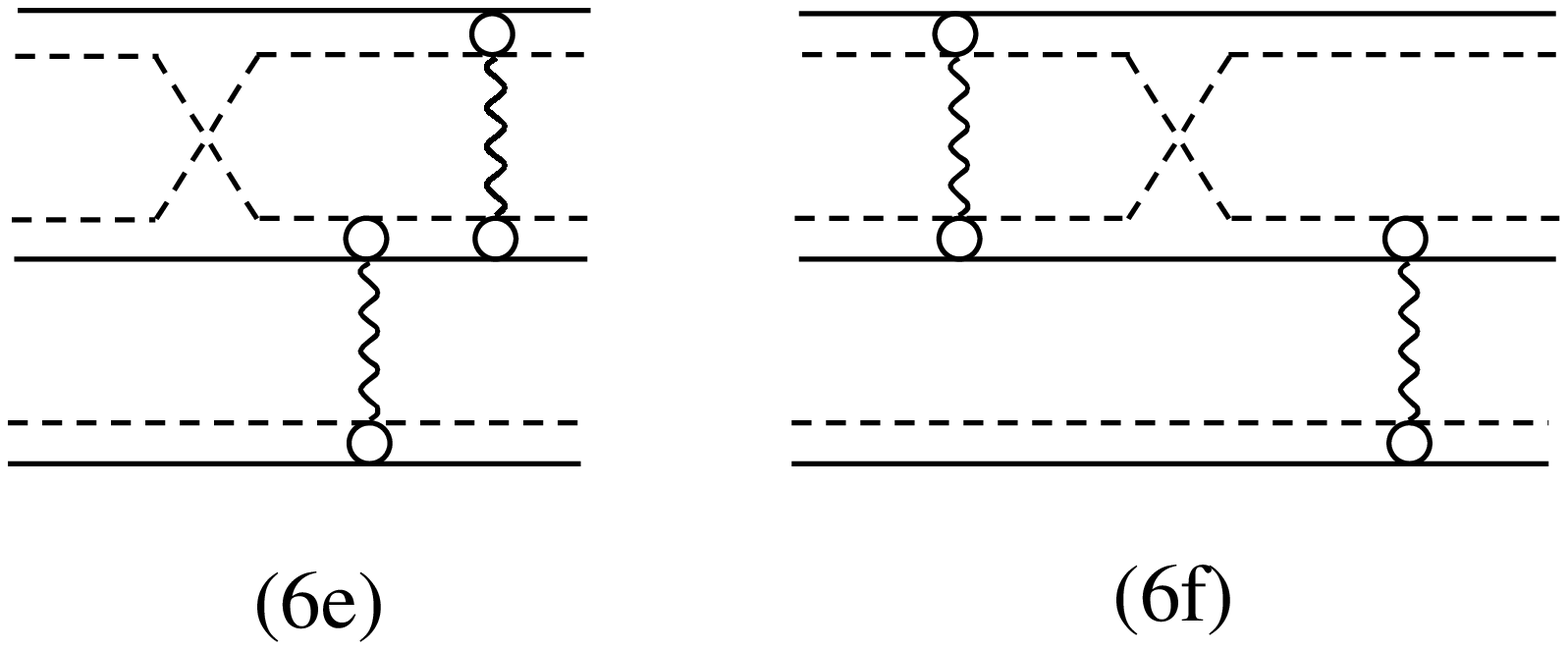}}}
\end{figure}

\clearpage

\begin{figure}
\centerline{ \scalebox{0.7}{\includegraphics{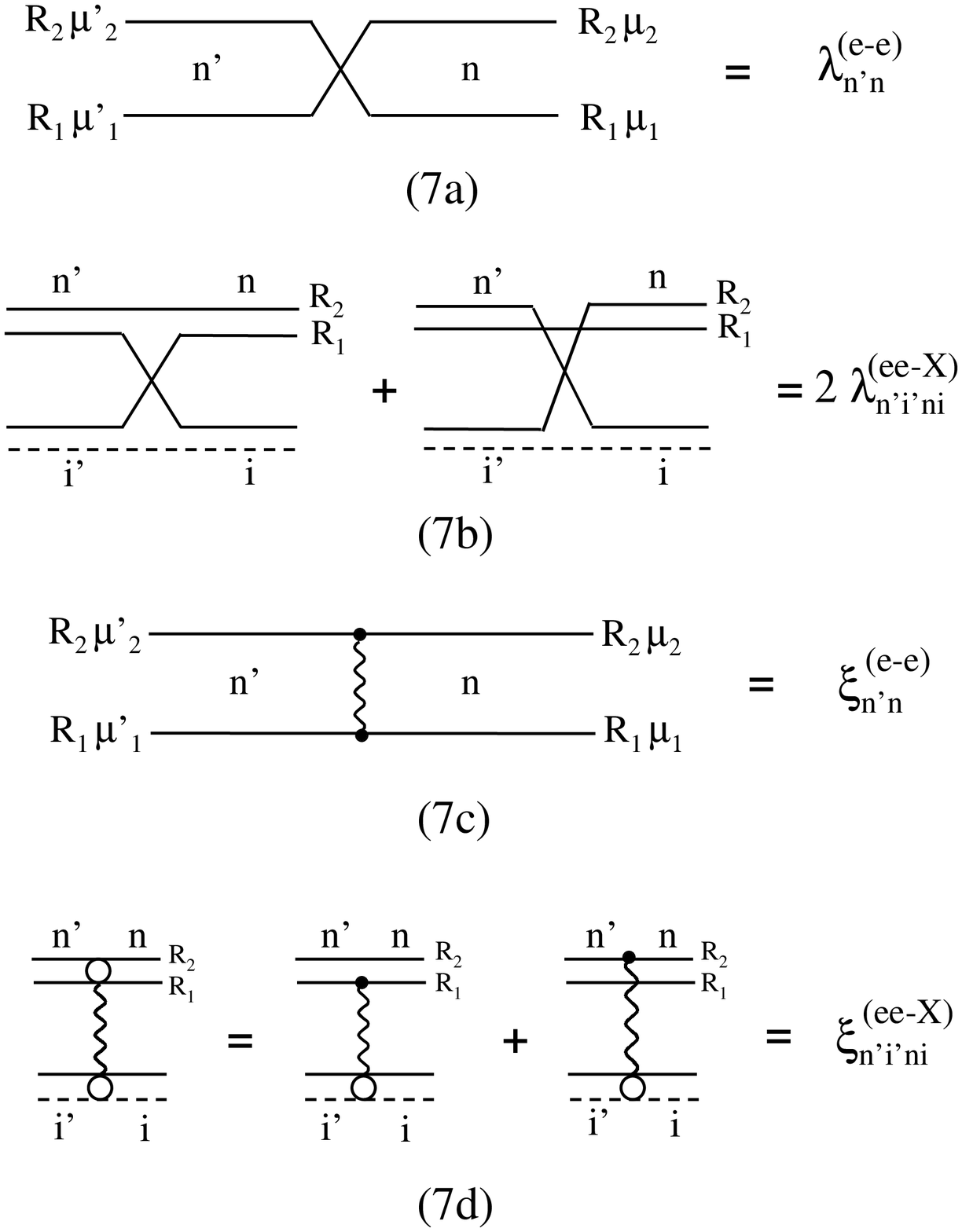}}}
\end{figure}

\end{document}